\documentclass[conference]{IEEEtran}
\IEEEoverridecommandlockouts
\usepackage{devanagari} % working
\usepackage{cite}
\usepackage{amsmath,amssymb,amsfonts}
\usepackage{algorithmic}
\usepackage{textcomp}
\usepackage[bottom]{footmisc}
\usepackage[utf8]{inputenc} % allow utf-8 input
\usepackage[T1]{fontenc}    % use 8-bit T1 fonts
\usepackage{hyperref}       % hyperlinks
\usepackage{url}            % simple URL typesetting
\usepackage{booktabs}       % professional-quality tables
\usepackage{nicefrac}       % compact symbols for 1/2, etc.
\usepackage{microtype}      % microtypography
\usepackage{lipsum}
\usepackage{graphicx}
\usepackage{subfig}
\usepackage{tabularx}
\usepackage{multirow}

\newcolumntype{P}[1]{>{\centering\arraybackslash}p{#1}}

\usepackage{fancyvrb}

\RecustomVerbatimCommand{\VerbatimInput}{VerbatimInput}%
{fontsize=\footnotesize,
 framesep=0em, % separation between frame and text
 rulecolor=\color{Gray},
 label=\fbox{\color{Black}data.txt},
 labelposition=topline,
 commandchars=\|\(\), % escape character and argument delimiters for
 commentchar=*        % comment character
}

\def\BibTeX{{\rm B\kern-.05em{\sc i\kern-.025em b}\kern-.08em
    T\kern-.1667em\lower.7ex\hbox{E}\kern-.125emX}}
\begin{document}

\title{What's Kooking? Characterizing India's Emerging Social Network, Koo}

\author{
    \IEEEauthorblockN{
        Asmit Kumar Singh\IEEEauthorrefmark{1}\textsuperscript{\textsection},
        Chirag Jain\IEEEauthorrefmark{1}\textsuperscript{\textsection},
        Jivitesh Jain\IEEEauthorrefmark{2}\textsuperscript{\textsection},
        Rishi Raj Jain\IEEEauthorrefmark{1}\textsuperscript{\textsection},\\
        Shradha Sehgal\IEEEauthorrefmark{2}\textsuperscript{\textsection},
        Tanisha Pandey\IEEEauthorrefmark{1}\textsuperscript{\textsection},
        Ponnurangam Kumaraguru\IEEEauthorrefmark{2}\thanks{\# Major part of this work was done while Ponnurangam Kumaraguru was a faculty at IIIT-Delhi.}\textsuperscript{\#}
    }
    \IEEEauthorblockA{\IEEEauthorrefmark{1}Indraprastha Institute of Information Technology, Delhi, India}
    \IEEEauthorblockA{\IEEEauthorrefmark{2}International Institute of Information Technology, Hyderabad, India}
    \IEEEauthorblockA{\{asmit18025, chirag17041, rishi18304, tanisha17116\}@iiitd.ac.in}
    \IEEEauthorblockA{\{jivitesh.jain, shradha.sehgal\}@students.iiit.ac.in, pk.guru@iiit.ac.in}
}

\maketitle

\begingroup\renewcommand\thefootnote{\textsection}
\begin{NoHyper}
\footnotetext{Equal contribution. Arranged in the alphabetical order of the first name.}
\end{NoHyper}
\endgroup

\begin{abstract}
Social media has grown exponentially in a short period, coming to the forefront of communications and online interactions. Despite their rapid growth, social media platforms have been unable to scale to different languages globally and remain inaccessible to many. In this paper, we characterize Koo, a multilingual micro-blogging site that rose in popularity in 2021, as an Indian alternative to Twitter. We collected a dataset of 4.07 million users, 163.12 million follower-following relationships, and their content and activity across 12 languages. We study the user demographic along the lines of language, location, gender, and profession. The prominent presence of Indian languages in the discourse on Koo indicates the platform's success in promoting regional languages.  We observe Koo's follower-following network to be much denser than Twitter's, comprising of closely-knit linguistic communities. An N-gram analysis of posts on Koo shows a \#KooVsTwitter rhetoric, revealing the debate comparing the two platforms. Our characterization highlights the dynamics of the multilingual social network and its diverse Indian user base. 
\end{abstract}

\begin{IEEEkeywords}
Social Computing, Data Mining, Online Social Media, Computational Social Science, Koo App, Twitter
\end{IEEEkeywords}

\section{Introduction}
\label{sec:intro}

With 4.66 billion users worldwide,\footnote{\url{https://bit.ly/3d2VzRB} Accessed 11 March 2021.} the Internet has become a mainstream medium for social interaction and information dissemination. Social media is an integral part of the Internet, as it enables users and organizations worldwide to connect, socialize, and express themselves with ease to a large audience. With 560 million Internet users, India ranks second in the world. However, English is the first language for only 0.02\% of the Indian population,\footnote{\url{https://bit.ly/3vtwWno} Accessed 11 March 2021.} thereby creating a language barrier for the majority of the population~\cite{e_gov_eng}. The diverse set of languages spoken in India raises the need for multilingual social media platforms in Indian languages.
\par
Recognising this need, Koo is a recent attempt at building a multilingual Indian social network. Formerly known as \textit{Ku Koo Ku}, it was launched in March 2020 as an Indian alternative to the popular social networking service, Twitter. Originally available in Kannada, Koo allows for and encourages discourse in Indian languages and currently supports 9 other Indian languages apart from English.\footnote{At the time of data collection, support for three of these was still under development.} It plans to include more Indian languages soon. 
\par
Koo gained popularity during August 2020, when it won the Government of India's \textit{Aatmanirbhar} Bharat App Innovation Challenge Award.\footnote{\url{https://bit.ly/2RWGI3w} Accessed 12 March 2021.} During the Indian Farmers' protest in January 2021, Twitter entered a week-long standoff with the Indian Government, over their refusal to block accounts that the Indian Government claimed were spreading misinformation.\footnote{\url{https://bit.ly/3zwWaUS} Accessed 21 March 2021.} Consequently, several Indian Government ministers,\footnote{\url{https://bit.ly/3gyWWZ7} Accessed 21 March 2021.} officials and agencies\footnote{\url{https://bit.ly/3xuECY1} Accessed 21 March 2021.} created accounts on Koo. There has been a general promotion of Koo amongst Government organizations since this event, sparking a rise in the platform's popularity. It again saw an increase in the influx of users when Koo became the first platform to agree to abide by the Government of India's Information Technology (Intermediary Guidelines and Digital Media Ethics Code) 2021 \cite{meity}, while Twitter and Facebook resisted them.
Koo has even announced its entry into Nigeria, with the Nigerian Government creating an account on the platform, following the ban of Twitter in the country.\footnote{\url{https://bit.ly/2RXEaCi} Accessed 14 June 2021.}
\par
% \todo{tanisha}
% Koo uniquely positions itself as a popular alternative to mainstream online social networks like Twitter. The reasons for this are twofold: (1) In the Indian context, its multilingual support allows for more inclusivity. The language barrier is broken, and more people can now join social media and express and share their opinions in the language that they choose. Moreover, (2) with the increasing support from the government agencies, now in both India and Nigeria, it becomes a strong competitor to mainstream platforms. With the ever-increasing political discourse on social media, Koo becomes an important platform from a political standpoint.  
Koo uniquely positions itself as an alternative to mainstream online social networks like Twitter. A cardinal reason for its popularity in the Indian context is its multilingual support that allows for more inclusivity. The language barrier is broken, and more people can now join social media and express and share their opinions in the language that they choose. Additionally, Koo has received support from the Indian and Nigerian government agencies, contributing to its acclaim. With the ever-increasing political discourse on social media, Koo becomes an important platform from a political standpoint.  
\par
The rich diversity of Indian languages on the platform coupled with its sudden rise and steady expansion in popularity and political backing motivate the need to understand what goes on inside such a unique platform. To understand this, we provide, to the best of our knowledge, the first characterization of the Koo social network. We do this by investigating the following research questions:
\begin{enumerate}
    \item \textbf{RQ1:} What are the characteristics and demographic of Koo users? When did they join the platform?
    \item \textbf{RQ2:} What kind of content is posted on the platform? Which languages are most popular? 
    % \todo{shradha}
    \item \textbf{RQ3:} What are the network properties of Koo and how do they differ from Twitter's? What communities are formed on the platform?
\end{enumerate}
We were able to collect data for 4.07 million users out of the total 4.7 million\footnote{As of 16th March 2021, https://bit.ly/3d2VcGH} users and 163 million follower-following edges in our data (Section~\ref{sec:data_collection}). In Section~\ref{sec:rq1}, we analyse the user demographic - we found that even though the users on the platform are predominantly male, female reported users are more active, and have more average followers and average likes. We also found that the influx of users surged on the platform in August 2020 when it won the \textit{Atmanirbhar} Bharat App Innovation Challenge, and in the starting months of 2021 when the government promoted the App. On analysing the content in Section~\ref{sec:rq2}, we saw that the most popular language on Koo is Hindi, followed by English, Kannada and Telugu. We also observed a Koo vs Twitter rhetoric on the platform and support for the Indian political party BJP. We saw that major political figures were one of the most mentioned users on Koo. On comparing the network properties of Koo and Twitter (Section~\ref{sec:rq3}), we found Koo to have a dense, well connected network with a higher clustering coefficient. Furthermore, we observed distinct communities of users on Koo, which are based on languages, with English speaking users more centrally placed and having connections to Hindi and Kannada users. 
\par
Through our work, we make the following contributions:
\begin{enumerate}
    \item \textit{Perform an extensive characterization of the new Indian social network Koo, in terms of its user demographic and content.}
    \item \textit{Present the first dataset of users, their connections, and content on Koo.}
    \item \textit{Study the network and communities formed on this multilingual platform.}
\end{enumerate}

\section{Related Work}
In this section, we review previous work on social media analysis, in particular the multilingual nature of platforms.
\par
There has been a vast amount of research on popular social media, Twitter. This includes its characterization in its initial years \cite{twitter2010} and the documentation of its steady user growth \cite{twt_pop}. With the growth of the platforms, the amount of research increased as well, with new topics of interest emerging, such as fake news and misinformation \cite{fake_news} \cite{misinformation}, automated bots \cite{bots}, hate speech \cite{hatespeech} \cite{hate_2}, etc. 
Some works have explored Indian languages on Twitter, with many using it as a corpus of indic-NLP research \cite{twt_ind_nlp} \cite{twt_ind_nlp_2} or studying the topics of discourse in the Indian subcontinent \cite{indian_twt} \cite{indian_twt_2}. Our work differs substantially from these as we focus on a platform dedicated to promoting Indian languages, with a predominantly Indian user base. 
\par
There have been studies exploring the emergence of new social platforms with unique properties such as Whisper for its anonymity \cite{whisper}, TikTok for its short videos \cite{tiktok}, Twitch for mixed media \cite{twitch}, and so on. Some Twitter alternative alt-right platforms like Gab \cite{zannettou2018gab} \cite{lima2018inside} and Parler \cite{prabhu2021capitol} have also been analysed in the past. Other social media platforms in local languages have also been studied, like VKontakte in Russia \cite{kozitsin2020modeling} and Weibo in China \cite{gao2012comparative}. 
\par
Multiple studies look at the multilingual aspect of other social platforms as well like blogs \cite{hale2012net} and reviews \cite{hale2016user}; but most of these cases find the English language to be dominant over the other languages. Our study focuses on a platform that has a native language to be dominant. Agarwal et al. \cite{sharechat} do study the qualities of an Indian multilingual social network, Sharechat. However, it focuses mainly on image-based content posted by the user. Koo is majorly a text-based platform, thereby making an inquiry of language aspects even more pertinent. Apart from the content posted, our work also lays emphasis on the user characteristics and network of Koo and draw a comparison of the platform's properties to that of Twitter. The choice of a novel platform and the context behind its rapid rise in popularity differentiates our work from the remaining.

\section{Methodology}
\label{sec:data_collection}
We describe next our data collection methodology. We start by presenting an overview of the platform.

\subsection{Overview of Koo platform}
Similar to Twitter, Koo allows logged-in users to share microblog posts known as ``\textit{koo}s''. While signing up, users can choose their display name, handle, language and other personal details such as gender, marital status, and birth date. Once logged-in, users can post koos, which can at most be 400 characters long; whereas, Twitter  allows tweets to be 280 characters at maximum. Users can also comment on others' koos or re-share (also known as ``\textit{rekoo}'') them with or without adding a comment. Koo provides transliteration support while typing in native Indian scripts. The platform's user interface, the users' feed, and the list of recommended accounts to follow changes according to their chosen language, enabling the formation of linguistic communities on the platform. 

% The Koo mobile application is more feature rich than the web application at the time of this study. The web application does not yet support the languages, Marathi and Bangla, and lacks direct message functionality, which is only available on the mobile application. Moreover, users cannot add their location, or link other social media handles from the web application. Unlike the mobile application, the website does not have sanity checks for the date of birth field, allowing users to enter dates even in the future.

\subsection{Data Collection}

We collected data pertaining to the users' profiles, the follower-following network, and content (koos, rekoos, comments, likes, and mentions) posted on the platform, from Koo's public API that serves the Koo web and mobile applications. Data collection started on 26 February 2021 and lasted until 11 March 2021. Table~\ref{tab:data_stats} summarises statistics of our proposed dataset.

% We collected data pertaining to the users' profiles, the follower-following network, and content (koos, rekoos, comments, likes, and mentions) posted on the platform, from Koo's public API that serves the Koo web and mobile applications (see Appendix~\ref{app:endpoints} for the list of endpoints used). Data collection started on 26 February 2021 and lasted until 11 March 2021. Table~\ref{tab:data_stats} summarises statistics of our proposed dataset.

\par
Koo's stark resemblance to Twitter also invites a comparison between user behaviour on the two platforms. Hence, we created a dataset of Koo and Twitter user IDs that correspond to the same entity. This dataset can be used for automated identity resolution tasks and cross-platform analysis of Koo and Twitter.

\par
We make our dataset public in adherence to FAIR principles as described in Section~\ref{sec:fair_principles}.

\begin{table}[h!]
	\centering
	\begin{tabular}{lr}
		\toprule
		\textbf{Entity}                           & \textbf{Count}       \\
		\midrule
		User profiles                    & 4,061,735   \\
		Follower-following relationships & 163,117,465 \\
		Koos                             & 7,339,684   \\
		Rekoos                           & 2,828,158   \\
		Rekoo with Comments 
		                                 & 413,955     \\
		Comments                         & 4,793,492   \\
% 		Twitter user profiles            & 872         \\
		\bottomrule
	\end{tabular}
	\caption{Statistics of the dataset we collected from Koo.}
	\label{tab:data_stats}
\end{table}

\subsubsection{Koo Data Collection}

We collected and analyzed data about the general discourse on the platform and the indulgent users, without restricting ourselves to certain trends, hashtags, or topics. Analyzing close-to-complete networks enables a significant understanding of the entire platform~\cite{twitter_fallacy}. We used the follower-following network to discover and collect users recursively using the snowball methodology, essentially traversing the network graph in a breadth-first manner, similar to Kwak, et al.~\cite{twitter2010} and Zannettou, et al.~\cite{zannettou2018gab}. We seeded the search with Koo's official language accounts (Table~\ref{tab:lang_accounts}) which are the first accounts shown to a new user, and the list of popular accounts that Koo recommends users to follow.\footnote{\url{https://www.kooapp.com/people}} These were chosen because of their large number of followers. We then collected the (previously undiscovered) followers and followees of these accounts and repeated the process for their follower-followee network. This approach is bound to leave out singletons and isolated communities; however, we hypothesize that they would not constitute a large share of the active users on the platform. We preferred this over enumerating and querying the API for all possible user IDs because of the significantly larger number of API requests the latter entails, which may have overloaded Koo's servers. We then used the user IDs to collect the users' profiles and the content (koos, rekoos, comments, likes, and mentions) generated by them, by querying the appropriate endpoints.

\begin{table}[h!]
	\centering
	\begin{tabular}{c}
        \includegraphics[width=0.4\textwidth]{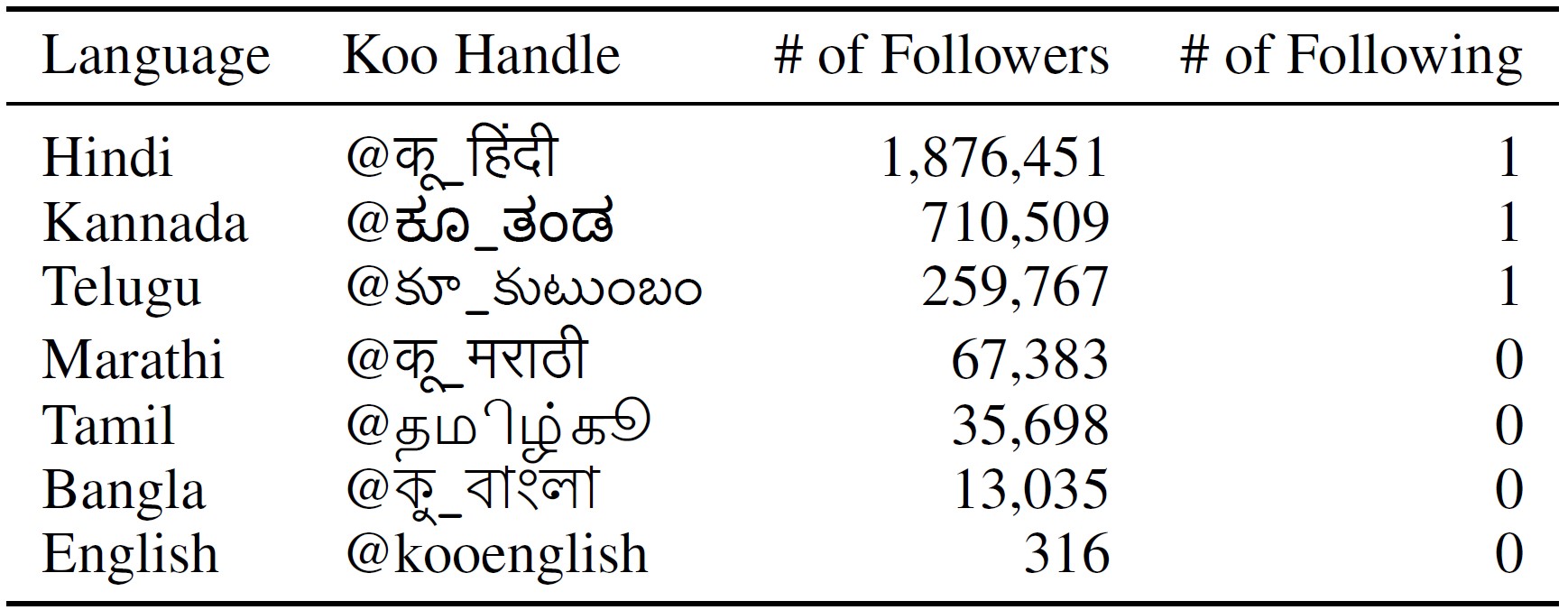}
    \end{tabular}
	\caption{Koo's official language accounts. Koo's Hindi account has the highest number of followers. Koo handles are written in the specific languages.}
	\label{tab:lang_accounts}
\end{table}

\subsubsection{Twitter Data Collection}
To create the Koo-Twitter user ID dataset, we considered two sets of users: those with verified Koo accounts and those who had listed their Twitter handles on their Koo profile. For the verified users, we manually curated corresponding Twitter handles. We queried Twitter's Users Search API\footnote{\url{https://api.twitter.com/1.1/users/search.json}} for the users' names and found probable matches, manually annotating each one. 1,030 verified Koo handles were distributed equally among 10 annotators, who were asked to verify the corresponding Twitter handles of the users. According to a set of guidelines prepared by the authors, the annotators looked for similarity in profile pictures, profile information, and the content the users posted on the two platforms. All the annotations were verified by two annotators and ambiguous cases were dropped from the dataset. Out of the 1,030 verified Koo users in our dataset, we found matching Twitter profiles for 872 of them, 499 of which are verified on Twitter as well.
\par
For the second case, we used Twitter handles provided by the users themselves. We removed duplicate handles and queried Twitter's Users Show API\footnote{\url{https://api.twitter.com/1.1/users/show.json}} to eliminate invalid usernames. In all, we make public 38,711 Koo and Twitter user IDs that correspond to the same entity.

% \textcolor{red}{
\subsection{Adherence to FAIR Dataset Principles}
\label{sec:fair_principles}
The gathered data consists of publicly available information about a social network, gathering and examining which would provide significant insights into the platform's characteristics. Our dataset also conforms to the FAIR principles. In particular, the dataset is ``findable'', as it is shared publicly.\footnote{\url{https://precog.iiit.ac.in/resources.html}} This dataset is also``accessible'', given the format used (CSV) is popular for data transfer and storage. This file format also makes the data ``interoperable'', given that most programming languages and softwares have libraries to process CSV files. Finally, the dataset is ``reusable'', as the included README file explains the data files in detail. The data was collected through public API endpoints of Koo, adhering to their privacy policy.\footnote{\url{https://www.kooapp.com/privacy}. Accessed 12 March 2021.} The data we collected was stored in a central server with restricted access and firewall protection. All experiments shown in this paper were performed on this dataset.
% }

% \section{Data Analysis}
\section{RQ1 : User Characteristics and Demographics}
\label{sec:rq1}
We analyze the demographics of users on Koo using their profile information. It should be noted that such user-entered information might not always be accurate and should be dealt with cautiously. 

\begin{figure*}[ht]
	\centering
	\includegraphics[width = 0.95\linewidth]{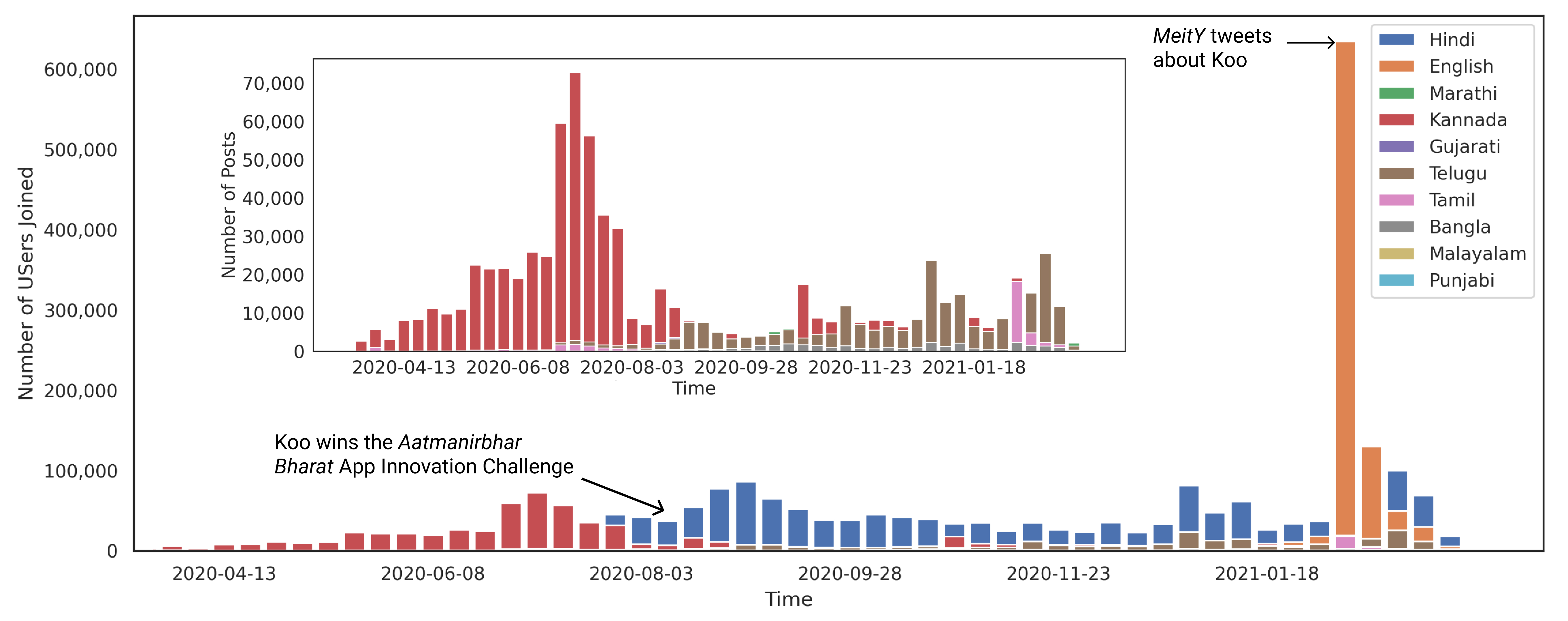}
	\caption{Weekly user creation timeline with language distribution, for 10 most popular languages. We observe peaks in users joining around August 2020, when Koo won the \textit{Aatmanirbhar} Bharat App Innovation Challenge, and 10 February 2021, when MeitY tweeted about Koo. Kannada language was popular during the initial stages of Koo, following which Hindi language took over from August 2020. Many English users started joining from February 2021.
	The inset graph represents the lower frequency language users (languages except for Hindi and English). }
	\label{fig:useron}
\end{figure*}

\subsection{User Onboarding} The platform is reported to host 4.7 million users at the time of writing. Our dataset has approximately 4 million users, of which, 1.9 million joined Koo in the first two months of 2021 alone. Figure~\ref{fig:useron} shows the user creation timeline across various languages. We find that Koo was predominantly used by Kannada users in its initial stages, presumably because it was launched in Bangalore where Kannada is the vernacular dialect. Surges in influx of users can be seen in August 2020 - around the time of the \textit{Aatmanirbhar} Bharat App Innovation Challenge Award. The user base expanded to other Indian languages, with Hindi users joining the platform in large numbers. February 2021 saw a huge spike in users with almost 200,000 of the users signing up on the days around 10 February 2021, just after the Ministry of Electronics and Information Technology, Government of India tweeted about the Koo app. As the debate of \#KooVsTwitter trended on Twitter, many English users joined the Koo platform. The inset graph of Figure~\ref{fig:useron} shows the distribution of languages other than Hindi and English - highlighting the prominence of Kannada, Telugu, and Tamil amongst Indian languages.

% \todo{shradha}
\subsection{Gender Distribution} 
\begin{figure*}[ht]
	\centering
	\begin{tabular}{cccc}
		\subfloat[Gender vs Content \label{fig:gender_act}]{\includegraphics[width = 0.22\linewidth]{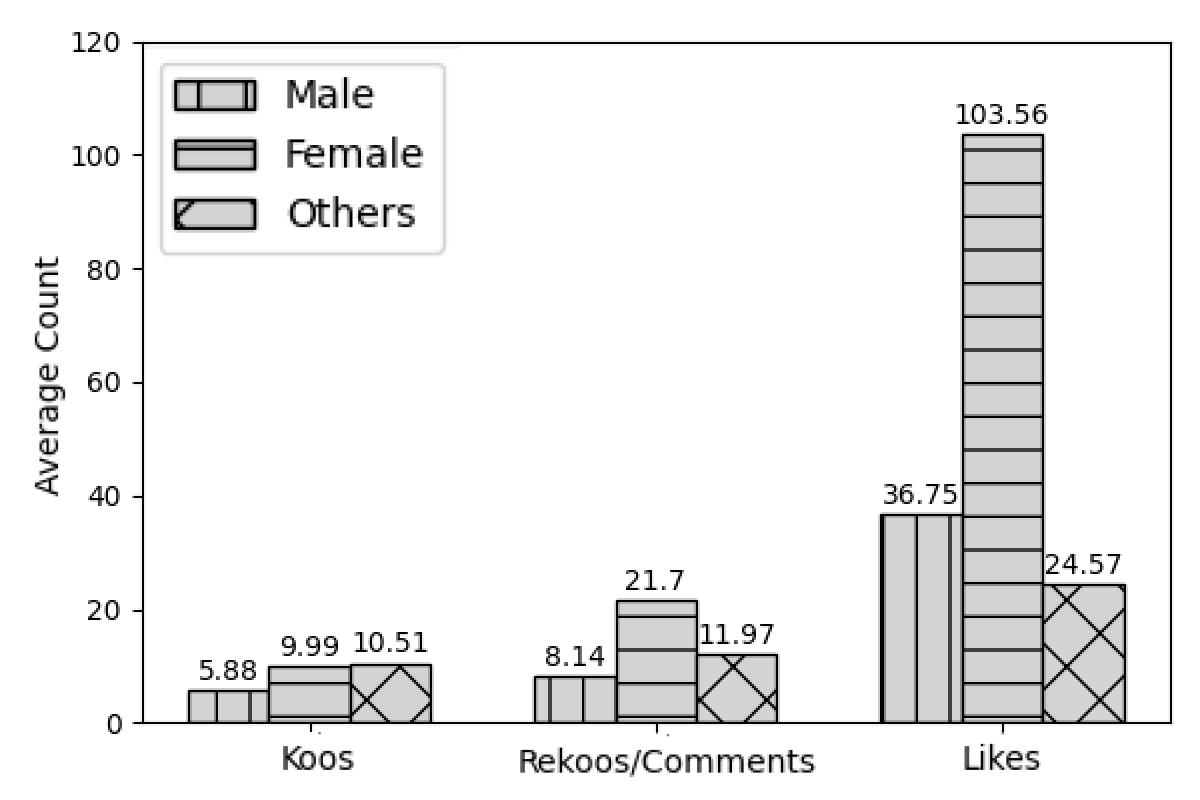}} & 
		\subfloat[Gender vs Follower \label{fig:gender_ff}]{\includegraphics[width = 0.22\linewidth]{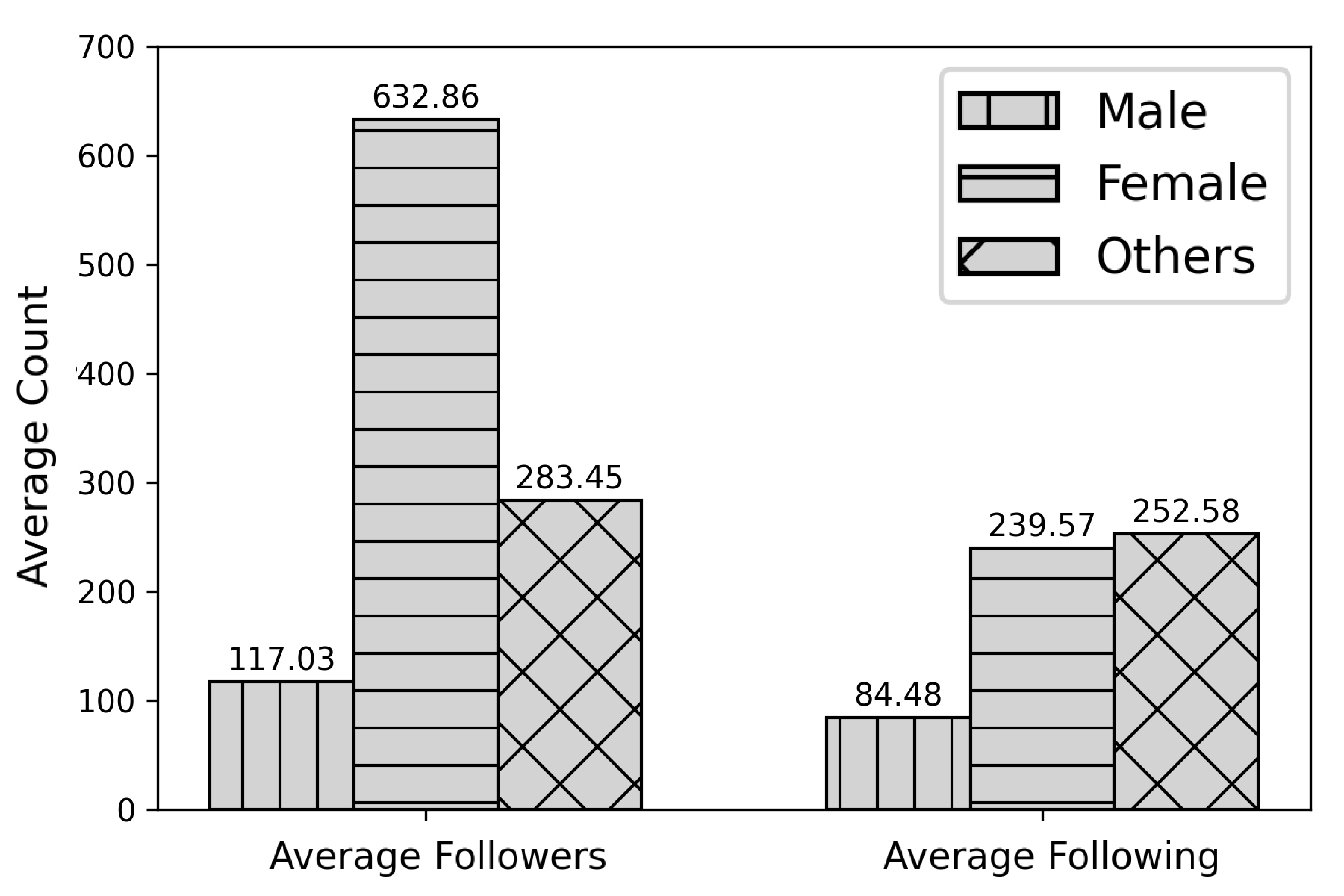}} &
 		\subfloat[Gender vs Age  \label{fig:gender_age}]{\includegraphics[width = 0.22\linewidth]{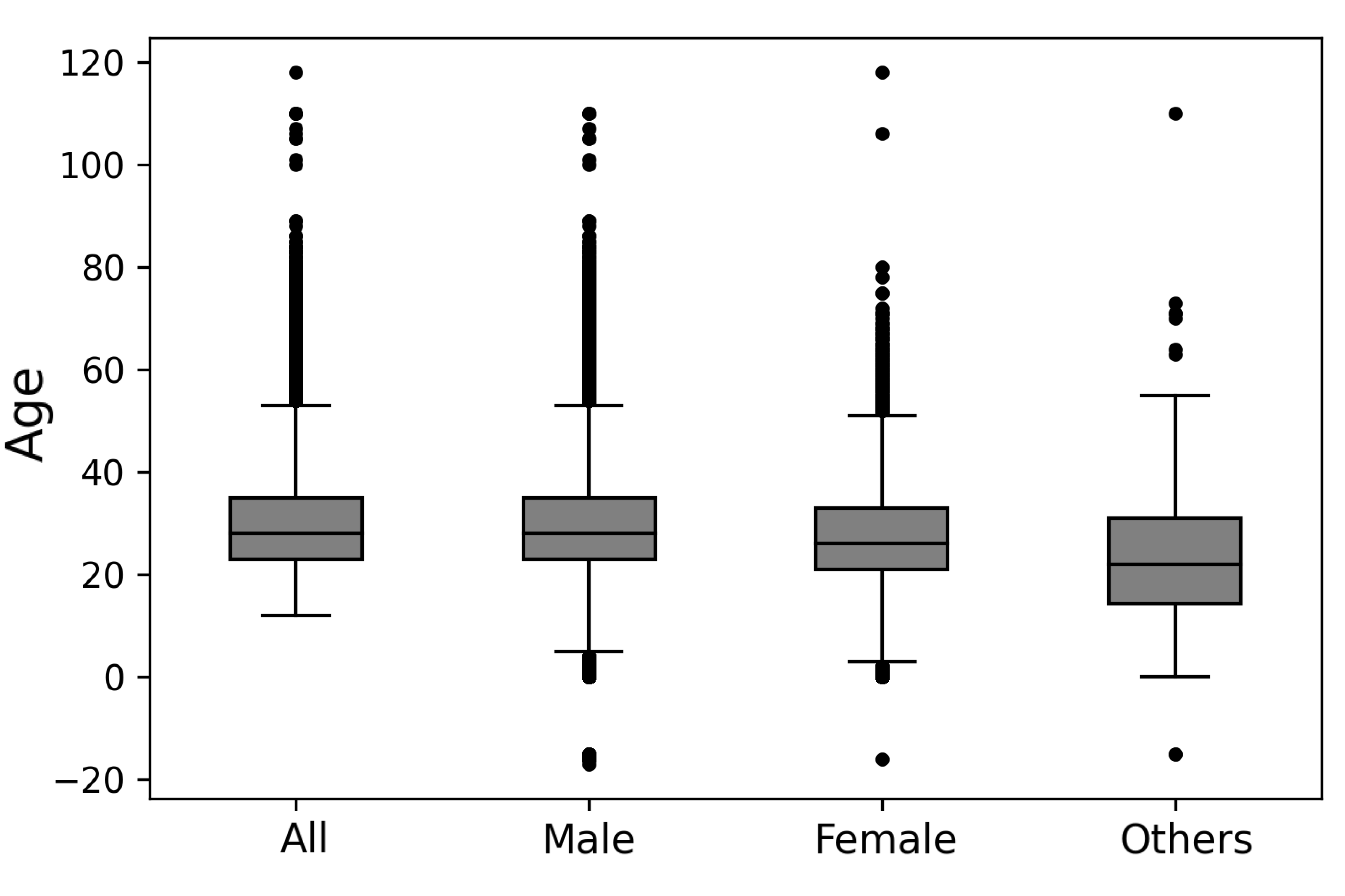}} &   
		\subfloat[Gender vs Marital status \label{fig:gender_marital}]{\includegraphics[width = 0.22\linewidth]{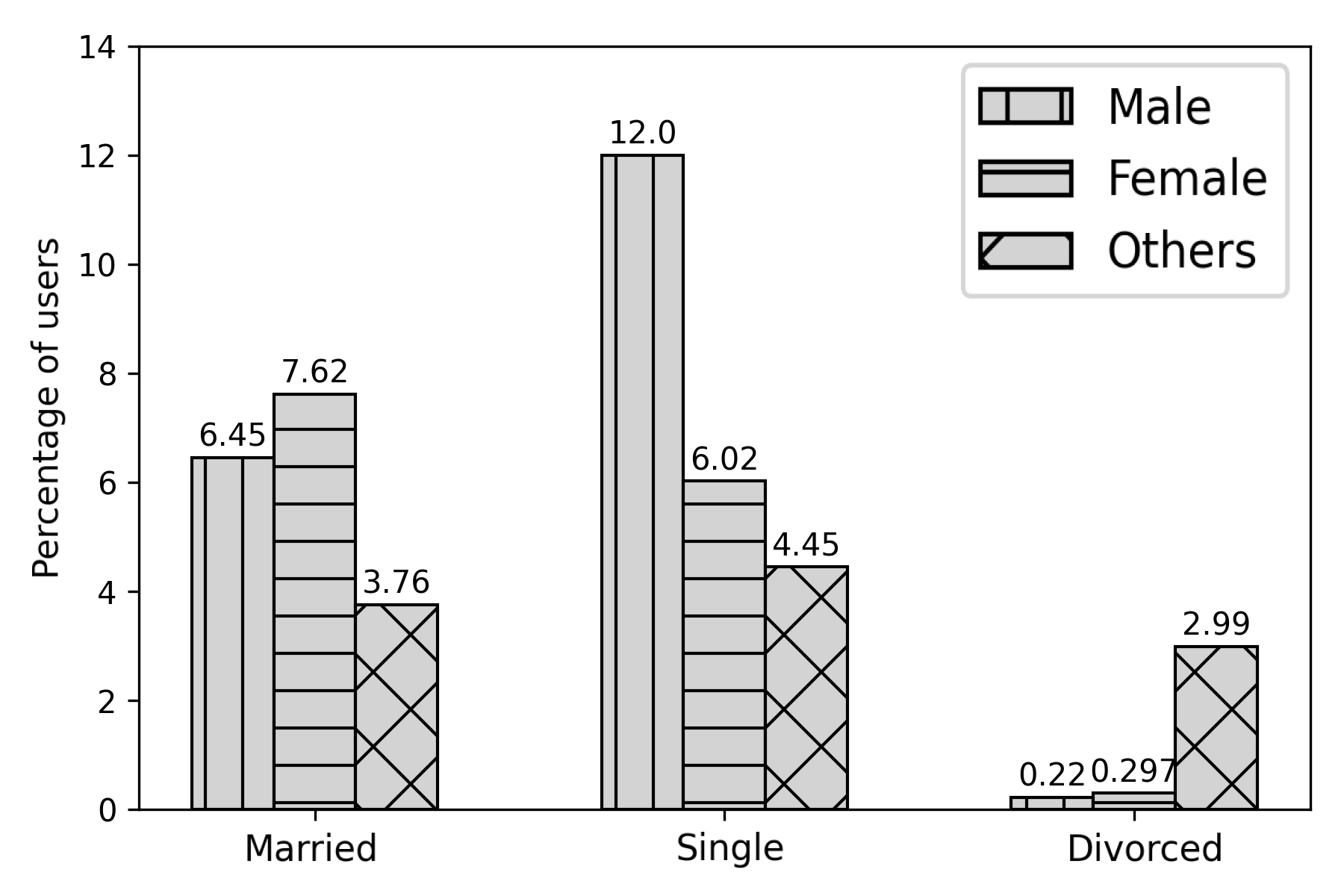}} 
						
	\end{tabular}
	\caption{Distribution of the user meta-data across genders. 18.1\% of total users specified their gender, $N = 761,315$, of which 92.1\% are male, 7.55\% are female and 0.362\% belong to the others category. (a) The box plot corresponds to the 25th, 50th and 75th quartile. The iOS App Store policy mentions a minimum age of 12. Hence, we discard all ages below 12 (1.2\%). Median age of 28 is observed across all users. (b) Female users have much higher average followers as compared to the other two categories (c) Female users are more active in terms of likes and rekoos as compared to the other two categories. (d) 12\% of male users are single which is much higher than the other two categories, while higher proportion of other category users are divorced.}
\end{figure*}

Of the 18.1\% of the users who specified their gender on their profile, 92.1\% identify as male (699,083), with only 7.5\% users identifying as female (58,996) and 0.36\% as others (3,236). However, Female users are more active, in terms of the number of average likes (103.6) and average rekoos (21.7) they do, as compared to other genders. Figure~\ref{fig:gender_act} also shows that Female have more followers (632.9) on average as compared to male users with an average of 117.0 followers and users identifying with the other category with an average of 283.45 followers. Male users, on average, produce fewer koos (5.9) and follow fewer people on average (84.8) than the other two categories, as visible in Figures~\ref{fig:gender_act} and \ref{fig:gender_ff}. Figure~\ref{fig:gender_age} shows that the median age of is similar for all gender categories at ~28 years. More male users identify themselves as single (12.0\%),  more female users as married (7.6\%) and more users of the other category as divorced (3.0\%), as shown in Figure~\ref{fig:gender_marital}. Table~\ref{tab:intergender}
shows the follower-following patterns between the genders. We observe that for all genders male users contribute to major proportion of the followers and following. 

% \begin{table}[h!]
% \centering
% \begin{tabular}{c c c c p{0pt} c c c}
% % \begin{tabular}{c c c c @{\hspace{1em}} c c c}
% \toprule
% \multirow{2}{*}{Gender}                  & \multicolumn{3}{c}{Followers}  && \multicolumn{3}{c}{Following}  \\ 
% \cline{2-4}
% \cline{6-8}
%                                  & Male & Female & Others && Male & Female & Others \\ 
% \midrule
% Male & 87.90\% & 11.14\% & 0.95\% && 75.61\% & 23.56\% & 0.81\% \\
% Female & 92.12\% & 7.32\% & 0.54\% && 83.19\% & 16.20\% & 0.59\% \\
% Others & 91.59\% & 7.84\% & 0.56\% && 84.85\% & 14.61\% & 0.52\% \\
% \bottomrule
% \end{tabular}
% 	\caption{Gender wise follower-following distribution. Male users contribute to the majority of followers and following across all the genders.}
% 	\label{tab:intergender}
% \end{table}

\begin{table}[h!]
	
	\centering
	\begin{tabularx}{0.5\textwidth}
	    {X P{0.045\textwidth}P{0.045\textwidth}P{0.045\textwidth}P{0.001mm}P{0.045\textwidth}P{0.045\textwidth}P{0.045\textwidth}}
		\hline
		\toprule
		\multicolumn{1}{c}{\multirow{2}{*}{Gender}}  & \multicolumn{3}{c}{Followers}  & & \multicolumn{3}{c}{Following}   \\
		\cline{2-4}\cline{6-8}
        \multicolumn{1}{c}{} & Male    & Female  & Others &  & Male    & Female  & Others  \\
% 		Gender & Male Followers & Female Followers & Others Followers & Male Following & Female Following & Others Following \\ 
		\midrule
		Male    & 87.90\%          & 11.14\%            & 0.95\%             && 75.61\%          & 23.56\%            & 0.81\%             \\ 
		Female  & 92.12\%          & 7.32\%             & 0.54\%             && 83.19\%          & 16.20\%            & 0.59\%             \\ 
		Others  & 91.59\%          & 7.84\%             & 0.56\%             && 84.85\%          & 14.61\%            & 0.52\%             \\ 
		\bottomrule
	\end{tabularx}
	\caption{Gender wise follower-following distribution. Male users contribute to the majority of followers and following across all the genders.}
	\label{tab:intergender}
\end{table}

\subsection{Language and Location} 
\begin{figure}[ht!]
	\centering
	\includegraphics[width=0.95\linewidth]{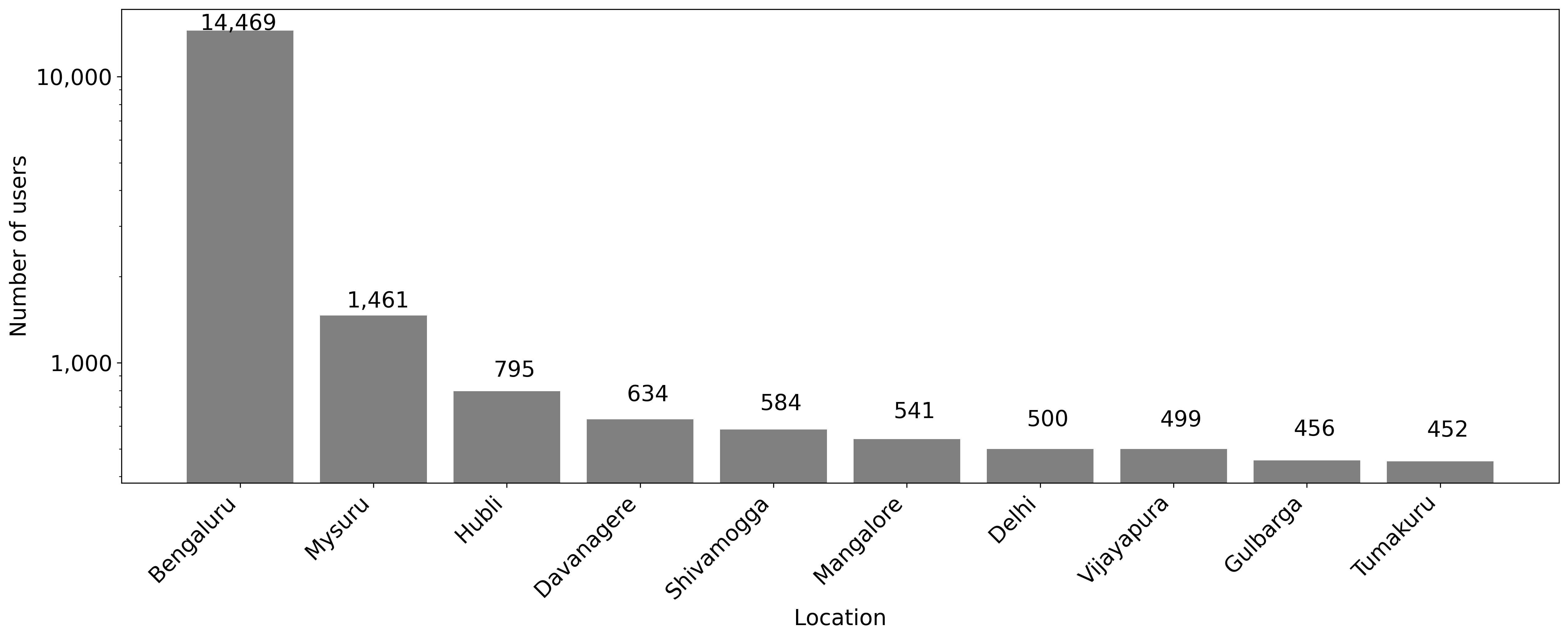}
	\caption{Top 10 current locations mentioned by users in the metadata, N=75,091. The y-axis is in logarithmic scale. Bengaluru is the most frequent location. Koo is headquartered in Bengaluru and was originally available in Kannada, which is Bengaluru's local language.}
	
	\label{fig:locations}
\end{figure}

Koo allows users to choose their location from a list of Indian cities. As shown in Figure~\ref{fig:locations}, in the 75,091 user profiles with location information, Bengaluru appears most frequently with 141,469 users, presumably because of the platform being headquartered in Bengaluru and being initially available in Kannada. Bengaluru also appears as an example help text over the location field on both the Android and iOS mobile applications.
\par

Despite Kannada being the first language on the platform, it is not the most popular, as it is outnumbered by Hindi, which was the language for 44.2\% and 51.2\% of the user base and total posts, respectively. Hindi was followed by English, which constituted to 23.8 \% of the users and 25.9\% of content (see Table~\ref{table:language}). The popularity of Hindi over English demonstrates a degree of success of the platform in promoting discourse in Indian languages. The distribution of users across languages closely mirrors the number of followers of Koo's official language accounts, apart from English (see Table~\ref{tab:lang_accounts}). This may be because the language account corresponding to the user's language is the first account whose post appears in a user's feed.

\subsection{Bio and Professional details} 
\begin{figure}[ht!]
	\centering
	\setlength{\fboxsep}{0pt}
		\setlength{\fboxrule}{1pt}
	\fbox{\includegraphics[width=.7\linewidth]{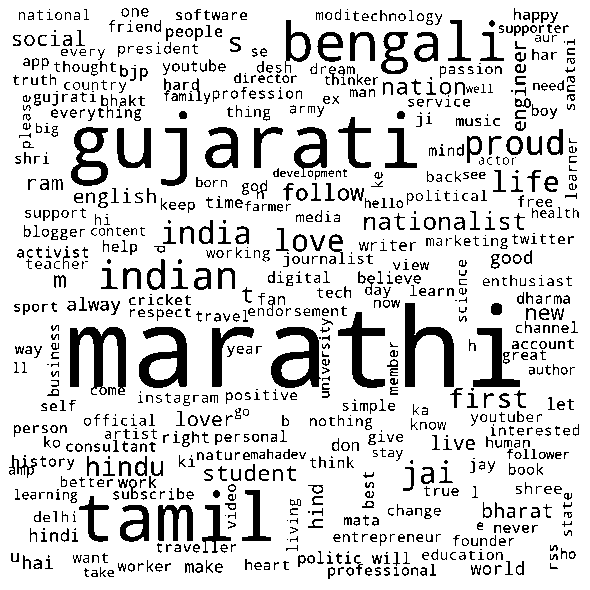}}
	\caption{Wordcloud for user bios. Presence of words pertaining to state and national identities such as ``Tamil'', ``Marathi'', ``Gujarati'' and ``Indian'' can be seen.}
	\label{fig:userbio}
\end{figure}

\begin{figure*}[ht!]
	\centering
	\begin{tabular}{cc}
		\subfloat[Work Title.]{\includegraphics[width = 0.4\linewidth]{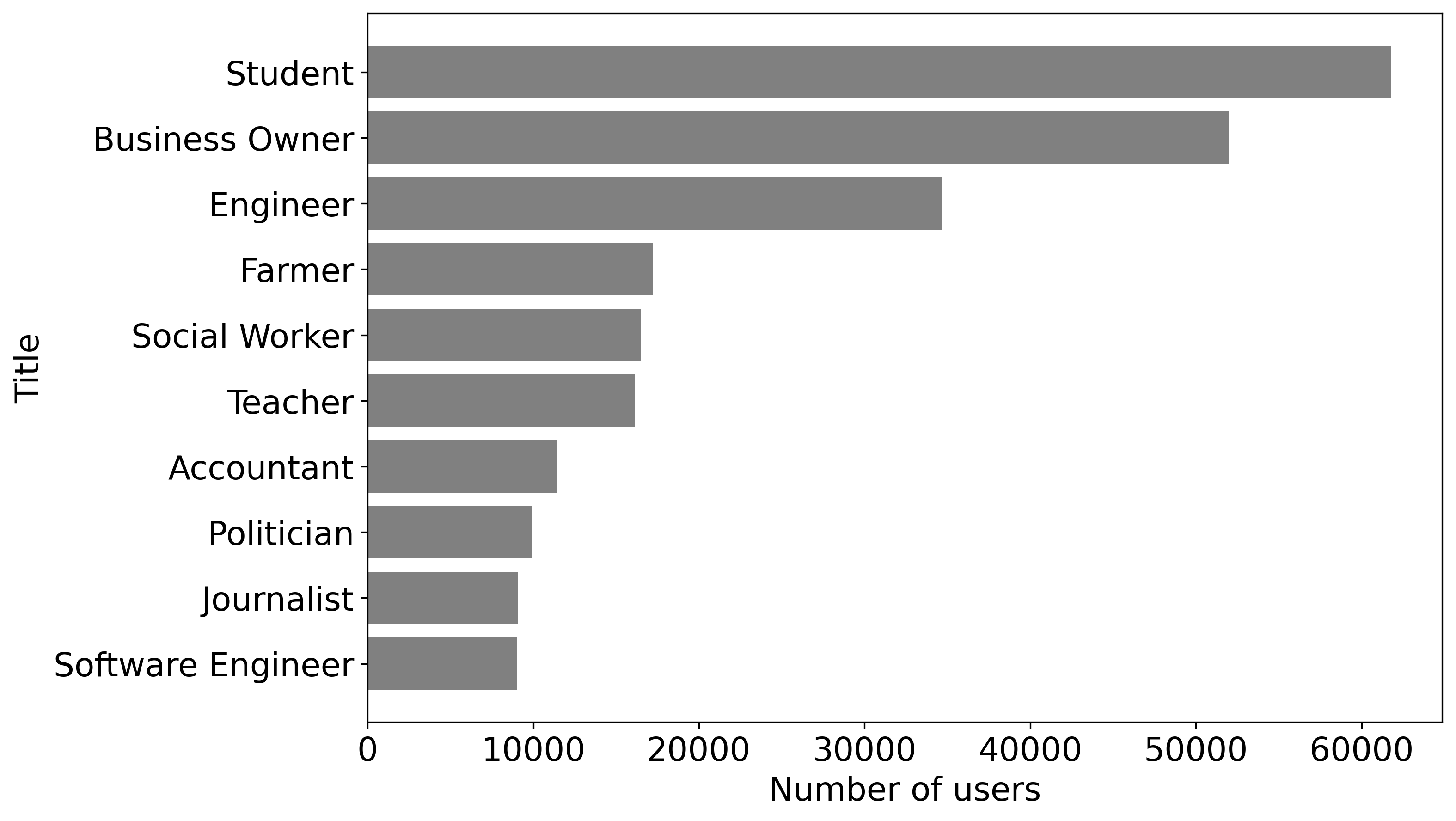}} &
		
		\subfloat[Educational Qualifications.]{\includegraphics[width = 0.4\linewidth]{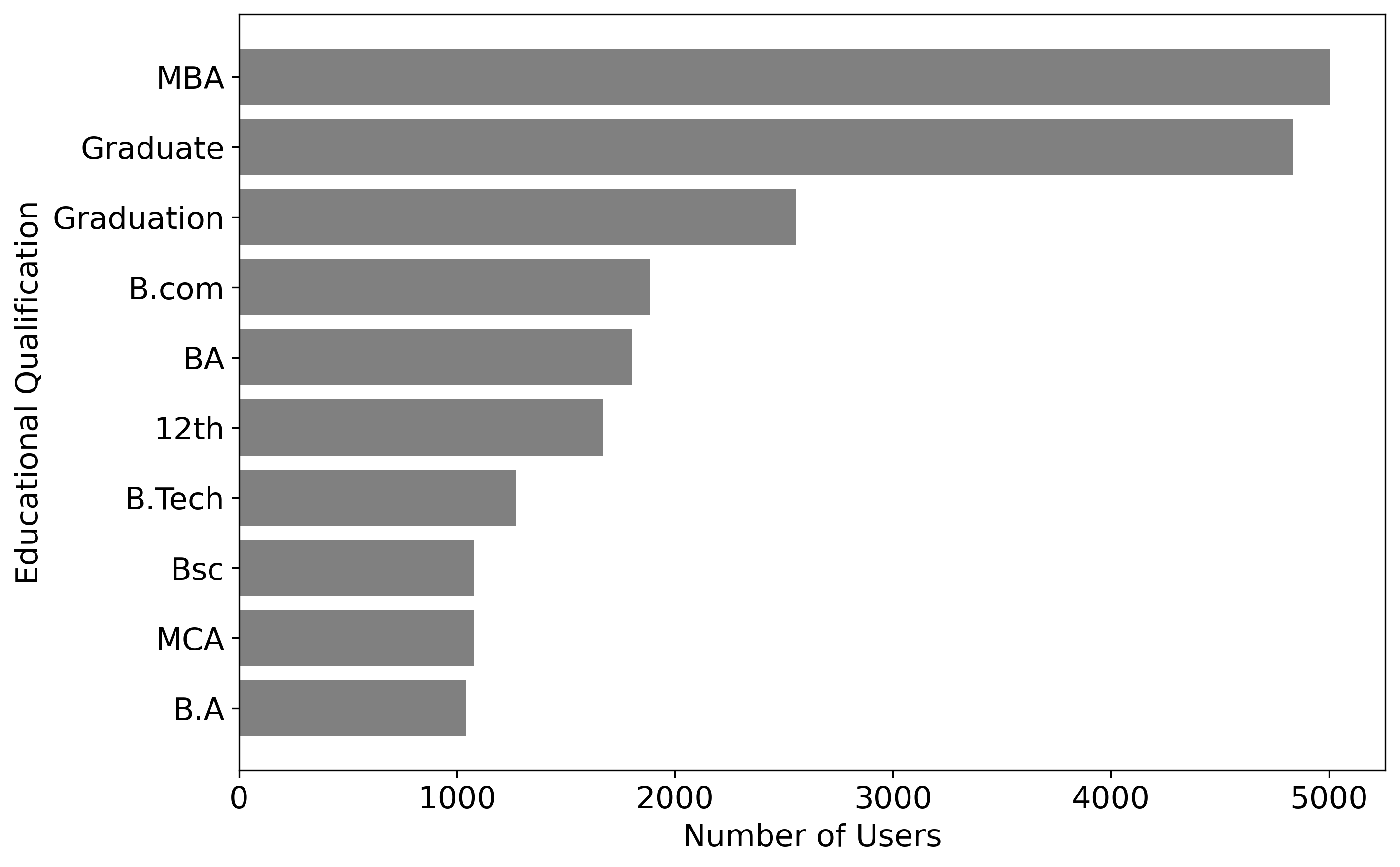}} 
	\end{tabular}
    	\caption{Top 10 occurring work titles and educational qualifications. (a) A total of N =  584,352 users mentioned their title, with ``student'' being the highest mentioned title with 61,778 occurrences (b) A total of N = 118,474 users mentioned educational qualifications, with ``MBA'' being the most common qualification with 4,988 mentions. }
	\label{fig:professional}
\end{figure*}

Koo allows the users to add and edit their profile bio. Figure~\ref{fig:userbio} shows the wordcloud of the user profile bios on Koo. We see the occurrence of words pertaining to state and national identities like ``Marathi'', ``Tamil'', ``Bengali'', ``Indian'', ``Bharat'' etc. Users are also allowed to add and edit professional details. Figure~\ref{fig:professional} shows the top occurring work titles and educational qualifications of the users, with ``Student'' being the most common work title (61,778 users) and ``MBA'' being the most common educational qualification (4,988 users).

\begin{table*}[h!]
	\label{tab:lang}
	\centering
	\begin{tabular}{lrrrr}
		\toprule
		\textbf{Language} & \textbf{Number of Users} & \textbf{Percentage of Users} & \textbf{Number of Posts} & \textbf{Percentage of Posts} \\ 
		\midrule
		Hindi             & 1,795,411                & 44.2030                      & 3,755,829                & 51.1715                      \\ 
		English           & 968,271                  & 23.8388                      & 1,907,993                & 25.9955                      \\
		Kannada           & 711,049                  & 17.5060                      & 818,679                  & 11.1541                      \\ 
		Telugu            & 259,171                  & 6.3807                       & 359,874                  & 4.9031                       \\ 
		Marathi           & 183,073                  & 4.5072                       & 242,803                  & 3.3080                       \\ 
		Gujarati          & 64,829                   & 1.5960                       & 126,853                  & 1.7283                       \\
		Tamil             & 48,285                   & 1.1887                       & 77,981                   & 1.0624                       \\
		Bangla            & 31,211                   & 0.7684                       & 49,172                   & 0.6699                       \\ 
		Malayalam         & 318                      & 0.0078                       & 257                      & 0.0035                       \\
		Assamese          & 46                       & 0.0011                       & 113                      & 0.0015                       \\ 
		Punjabi           & 43                       & 0.0010                       & 43                       & 0.0005                       \\ 
		Oriya             & 27                       & 0.0006                       & 87                       & 0.0011                       \\ 
		\bottomrule
	\end{tabular}
	\caption{Language distribution on the platform. Hindi is by far the most popular language, for both user profiles as well as posting contributing to 44.20\% of the user profile languages and 51.17\% of the posts.}
	\label{table:language}
\end{table*}

\begin{figure*}[ht]
	\centering

    \includegraphics[width = 0.9\linewidth]{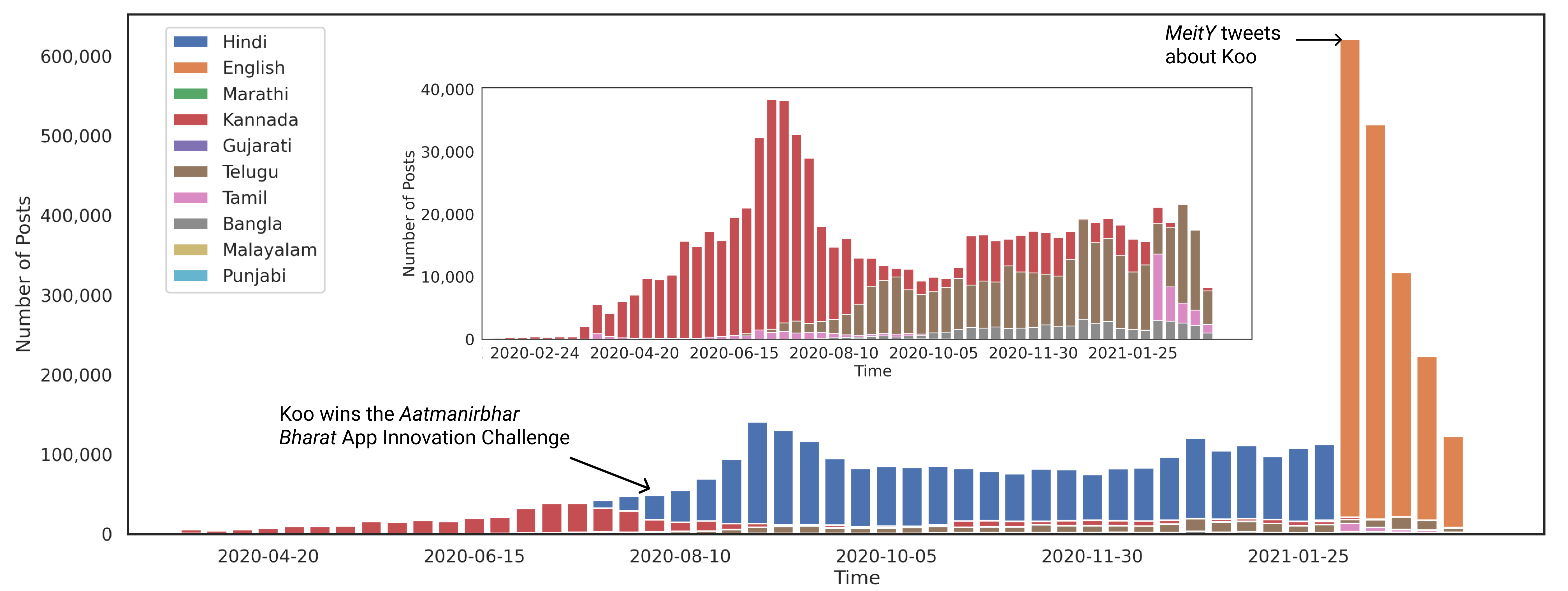}
    	\caption{Weekly post creation timelines with language distribution, for top 10 used languages. A peak around 10 February 2021 is observed, when MeitY promoted Koo on Twitter. Posts were majorly in Kannada language during the initial days of Koo, following which Hindi posts took over from August 2020. Posts in English language spiked around February 2021. The inset graph represents the lower frequency language posts (languages, except for English and Hindi). }
	\label{fig:posts}
\end{figure*}

\section{RQ2 : Content Analysis}
\label{sec:rq2}
\subsection{User-Generated Content} We have a total of 7,339,684 koos in our dataset. Figure~\ref{fig:posts} shows the post generation timeline of Koo with the language distribution. This displays a very similar behaviour to the user creation timeline. A major spike in posting is observed around 10 February 2021, especially in English koos, when the Government of India’s Ministry of Electronics and Information Technology posted a tweet promoting Koo, and many prominent government figures started joining the platform. The Kannada and Hindi language posts follow the same pattern as the user joining distribution. Similar to other social networks \cite{twitter_rhythm}, the posting activity on Koo gradually increases throughout the day, attaining its peak in the evening at around 2100 hours IST.

\subsection{Media Distribution}
\begin{figure}[!t]
	\centering
	
	\includegraphics[width=.85\linewidth]{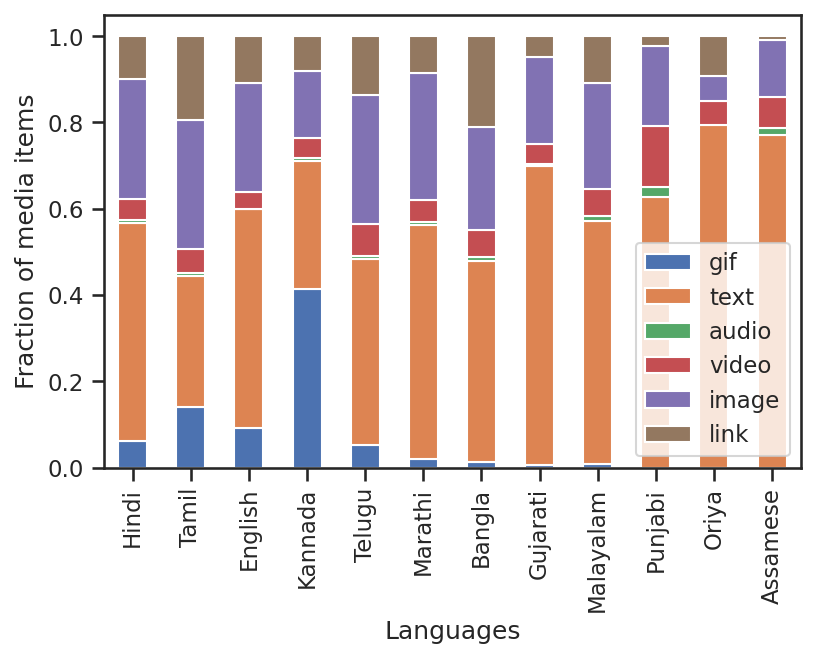}
	\caption{Media share in posts per language.}
	\label{fig:media}
\end{figure}

Koo allows for multiple types of media such as images, videos, links, gifs and text to be shared. Figure~\ref{fig:media} shows the distribution of these media types across various languages on the platform. Being a microblogging platform, more than 50\% of the content on the platform appears as text, followed by images, gifs, links, video and audio. Most languages follow the same general trend, with a stark exception of Kannada, showing 40\% of the content being shared through gifs.
\subsection{Hashtag Analysis}
\begin{figure}[!ht]
	\centering

			\includegraphics[width=\linewidth]{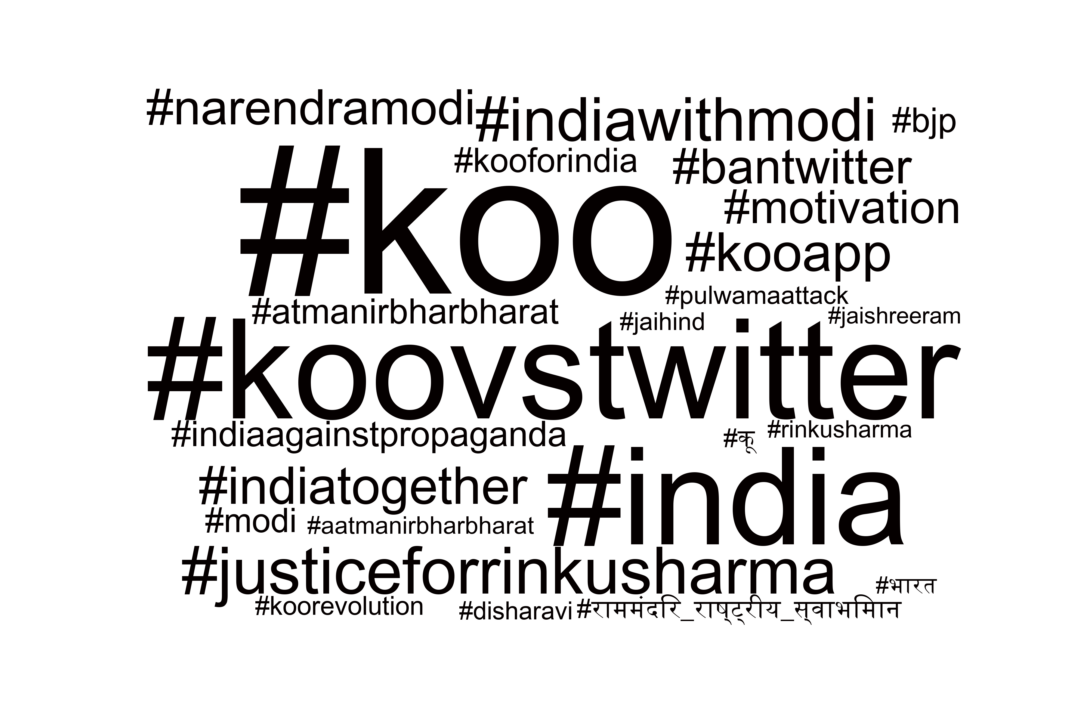}
		
		\caption{Wordcloud for top 25 hashtags in posts. Hashtags like \#koo, \#Koovstwitter, and \#bantwitter show the competitive sentiment between the two platforms.}
	\label{fig:hashtagwordcloud}
\end{figure}

\begin{figure}[ht!]
	\centering
	\includegraphics[width=\linewidth]{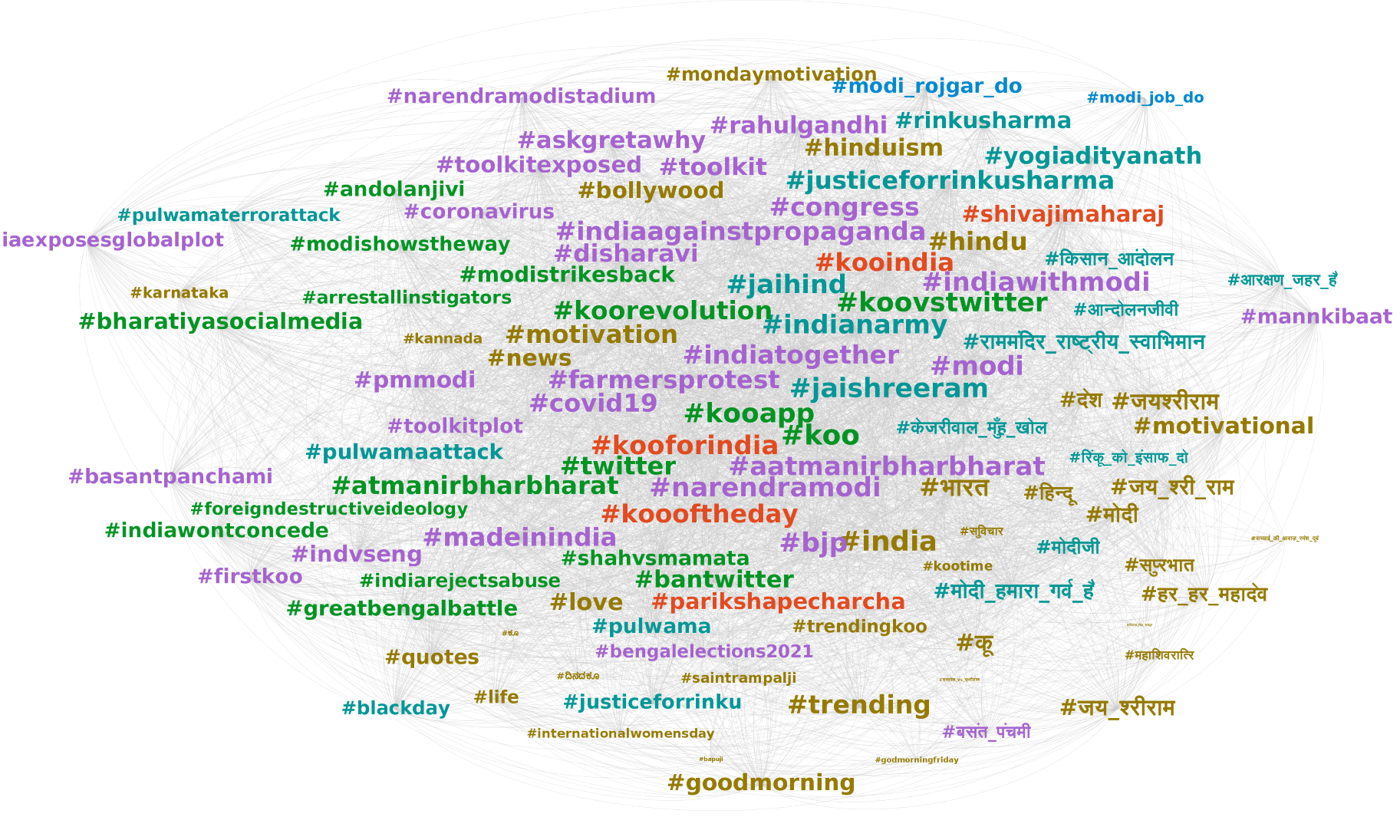}
	\caption{A network of the 100 most frequently occurring hashtags, where an edge represents co-occurrence of hashtags in a post. Colors indicate clustering based on modularity. The entire graph has a modularity score of 0.329.}
	\label{fig:hashtagclustering}
\end{figure}

Hashtags in social media have become a way for users to build communities around topics and promote opinions. We extract the top occurring hashtags from the user posts and plot them as a wordcloud in Figure~\ref{fig:hashtagwordcloud}. We see hashtags like ``kooforindia'', ``bantwitter'', and ``koovstwitter'', which project a sentiment of competition between Twitter and Koo, and promote the Koo platform. Hashtags like ``indiawithmodi'', ``atmanirbharbharat'', ``modi'', ``modistrikesback'', and ``bjp'', which are associated with  the Bharatiya Janata Party (BJP) are also present. 
Figure~\ref{fig:hashtagclustering} shows a network of the 100 most frequently occurring hashtags, with the edges indicating two hashtags that occur in the same post. Although the graph has a low modularity score (0.329), hashtags related to one topic fall in the same category when the graph is clustered based on modularity~\cite{Blondel_2008}.

\subsection{N-gram Analysis}
\begin{figure}[ht!]
	\centering
	\begin{tabular}{cc}
		\subfloat[Unigram Wordcloud.]
		{
		\setlength{\fboxsep}{0pt}
		\setlength{\fboxrule}{1pt}
		\fbox{\includegraphics[width = 0.4\linewidth]{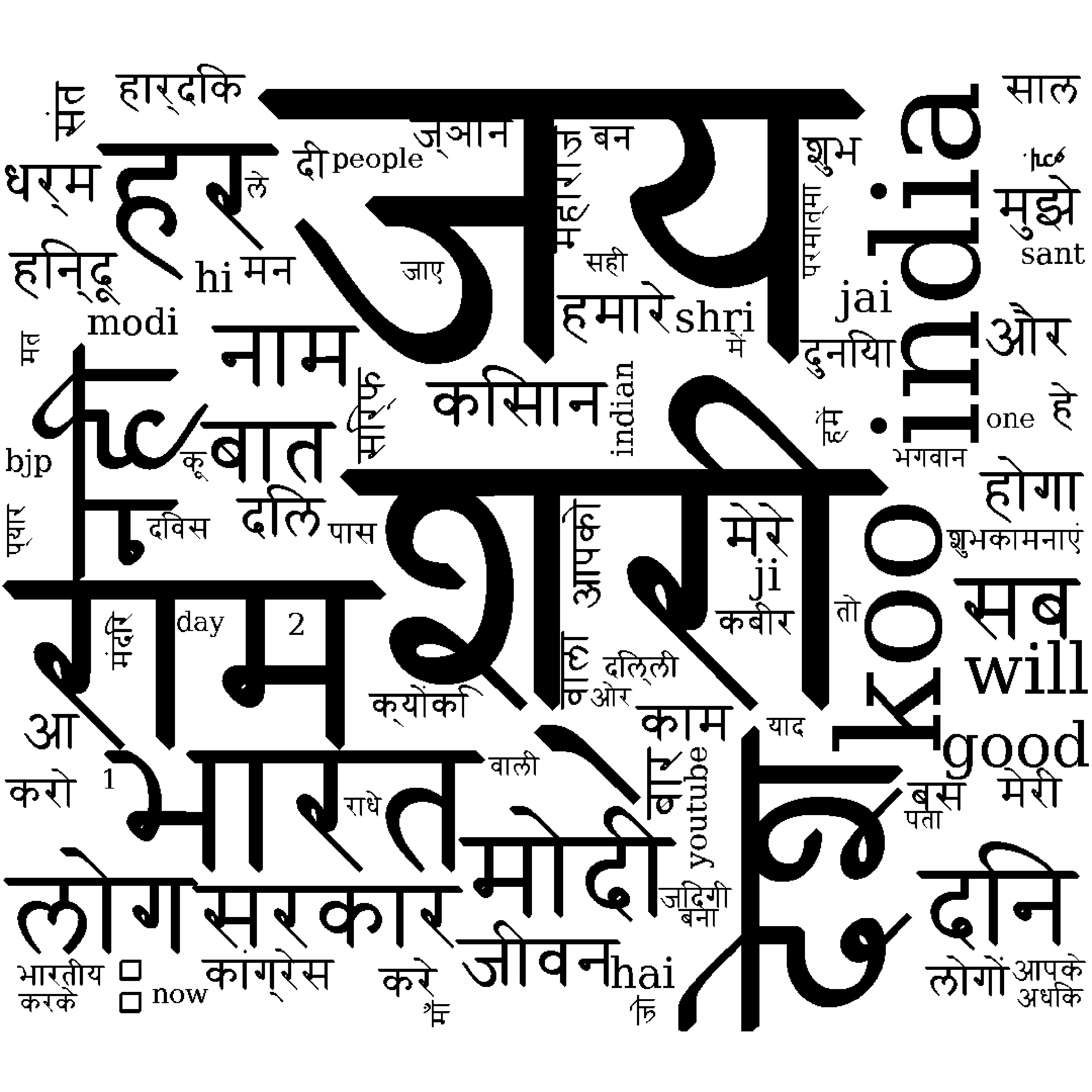}}
		}
		  &   
		\subfloat[Bigram Wordcloud.]
		{
		\setlength{\fboxsep}{0pt}
		\setlength{\fboxrule}{1pt}
		\fbox{\includegraphics[width = 0.4\linewidth]{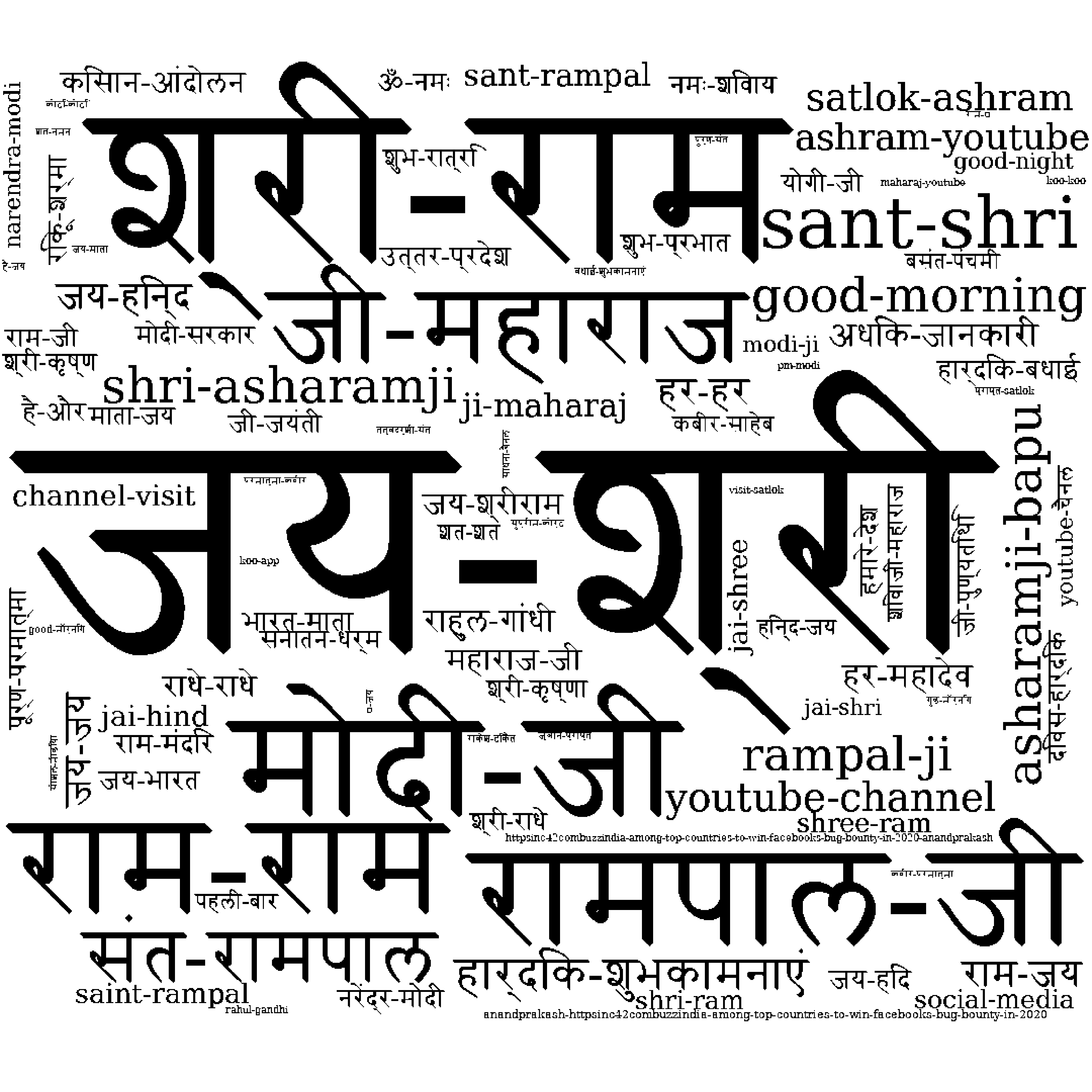}}
		} \\
	\end{tabular}
	\caption{Word Clouds for user posts. Both unigrams and bigrams show substantial Hindu religion centric content.}
	\label{ngrams}
\end{figure}

In order to get an insight into the popular conversations going on the platform, we plot the top occurring uni-grams and bi-grams in the content of the posts (Figure~\ref{ngrams}). We observe an overwhelming number of Hindi n-grams. Through both uni-grams and bi-grams, we observe the existence of Hindu-centric words like {\dn jy\399wFrAm} (``jaishreeram'') and {\dn rAm{\rs -\re}rAm} (``ram-ram'') that allude to ``Lord Ram'', a major deity in Hinduism. There is also a mention of many Indian religious leaders like {\dn rAmpAl{\rs -\re}jF} (``rampal-ji''), {\dn s\2t{\rs -\re}\399wF} (``sant-shri''), and {\dn jF{\rs -\re}mhArAj} (``ji-maharaj'')\footnote{Words in brackets are the transliterated forms of the original Hindi text.}. This indicates that some of the discussions on Koo are around religion and that people may be using slogans like {\dn jy\399wFrAm} (``jaishreeram'') as a symbol of their religious faith.

\subsection{Mentions and Likes}
The number of mentions and likes of a particular user are useful indicators to study a user's engagement and popularity on a platform~\cite{twitter_fallacy}. Table~\ref{tab:mentions} shows the top 10 most mentioned users on the platform, of which ``republic'', which is an Indian news channel comes at the top with 16,041 mentions. Notably, Republic also has an editorial partnership with Koo.\footnote{\url{https://bit.ly/35nZFPJ} Accessed 12 March 2021.} Ravi Shankar Prasad\footnote{Minister of Law and Justice, Electronics and Information Technology and Communications, Government of India.} is the next most mentioned user with 12,991 mentions. Many prominent ministers and political figures like Piyush Goyal\footnote{Minister of Railways, Commerce \& Industry, Consumer Affairs and Food \& Public Distribution, Government of India.} and Sambit Patra\footnote{Official spokesperson of the Bharatiya Janata Party in India.} are also present in the list of top mentioned users. Table~\ref{tab:likes} shows the top 10 users with the most liked posts on the platform. Ravi Shankar Prasad is the most liked user with 435,752 likes. Both English and Hindi pages for Republic News channel are present in the top 10 list.

\section{RQ3 : Koo's user network and communities}
\label{sec:rq3}
We analyzed Koo's following network, looking for characteristics of the community forming as the young platform grows. The network has a noticeably high local clustering coefficient of 0.561, that represents how well connected the neighbourhood of a vertex is. This indicates a strong modular structure in the network, presumably due to Koo only catering to audiences from a single country. In contrast, Twitter, which caters to worldwide audiences, only had an average local clustering coefficient of 0.072 during its early years in 2009~\cite{twitter_cc}, indicating much weaker communities.
\par

We see that around 90\% of the users have less than a hundred followers. However, a few users with an extremely high number of followers (of the order of $10^6$) skew the distribution. The distribution of the number of followees follows a similar pattern. Further, users with a high number of followers tend to only have a small number of followees and vice-versa, indicating the presence of influential and popular accounts that are followed by a majority of other users. This may, in part, be due to new users being shown a list of popular accounts to follow, on top of their feed.

\par
\begin{figure}[ht!]
	\centering
	\includegraphics[width=0.9\linewidth]{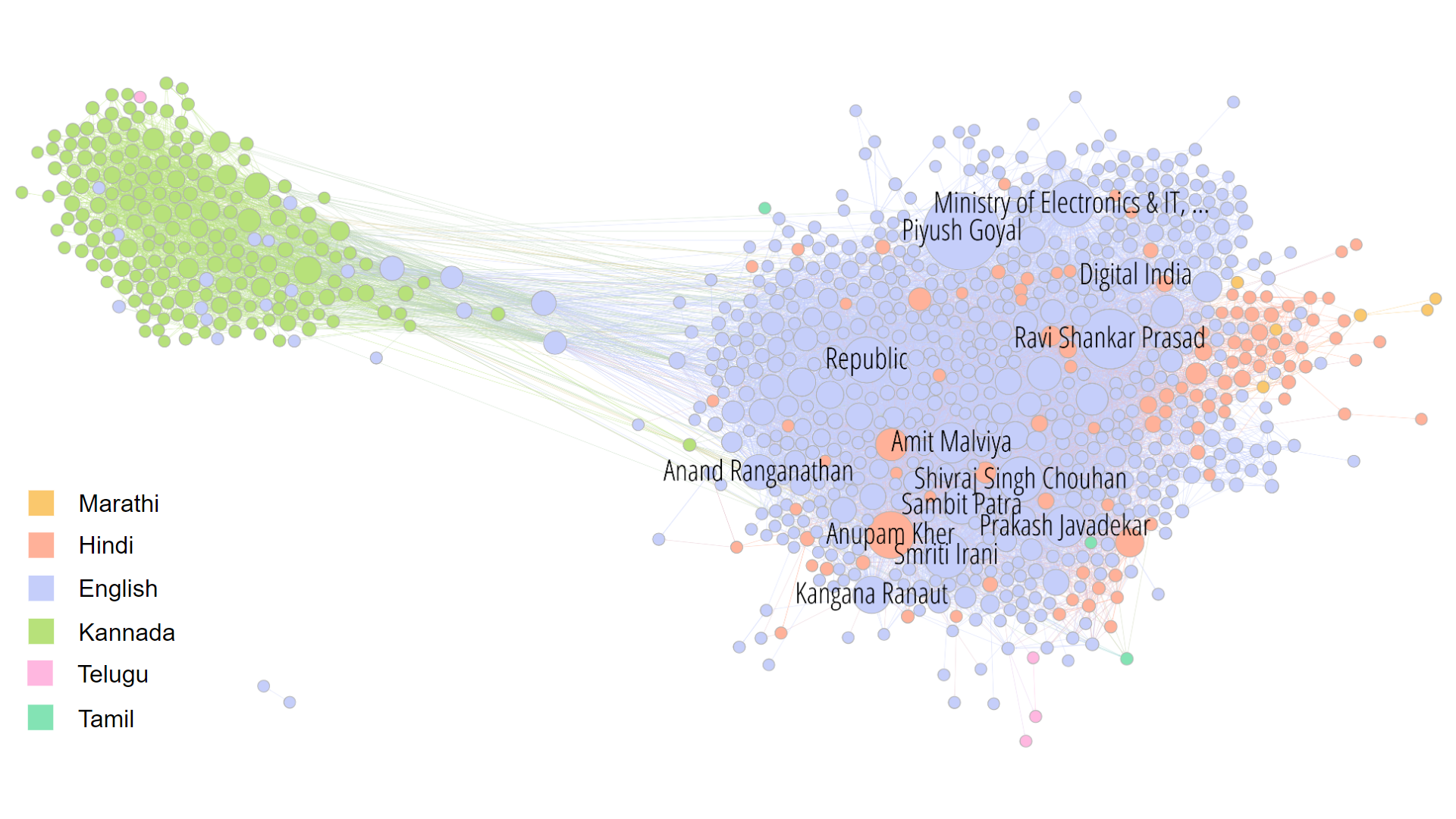}
	\caption{ Following graph of verified users on Koo. Edges are directed from the follower to the followee. Node size is proportional to in-degree, while colors indicate language. Names are only shown for a few prominent accounts. Singletons are not shown.}
	\label{fig:ff-verified}
\end{figure}

Figure~\ref{fig:ff-verified} shows the following network between verified users on Koo. We see distinct communities of users based on language, possibly because users interact more with others whose language they can understand. English-speaking users are more centrally placed in the network, with connections to both Hindi and Kannada speakers. Verified accounts in the two Indian languages, on the contrary, do not have many follower-followee relationships.

\begin{table}[ht!]
	\parbox{.48\linewidth}{
		\centering
		\begin{tabular}{lr}
			\hline
			\toprule
			\textbf{Handle}   & \textbf{Mentions} \\ 
			\midrule
			republic          & 16,041            \\ 
			ravishankarprasad & 12,991            \\ 
			kisanektamorcha   & 11,010            \\ 
			{\dn ErpENlk}\_{\dn BArt}     & 9,366     \\
			piyushgoyal       & 9,127             \\
			mayank            & 7,588             \\
			leledirect.com    & 7,045             \\
			aprameya          & 6,390             \\
			sambitpatra       & 5,742             \\
			khushbookapoor    & 5,693             \\
			\bottomrule
		
		\end{tabular}
		\caption{Users with most number of mentions. News channel Republic has the highest number of mentions (16,041), followed by Ravi Shankar Prasad (12,991).}
		\label{tab:mentions}
	}
	\hfill
	\parbox{.48\linewidth}{
		\centering
		\begin{tabular}{lr}
			\hline
			\toprule
			\textbf{Handle}         & \textbf{Likes} \\
			\midrule
			ravishankarprasad       & 435,752        \\
			piyushgoyal             & 395,674        \\
			republic                & 357,745        \\
            {\dn ErpENlk}\_{\dn BArt} & 296,039        \\
			meghupdates             & 277,495        \\ 
			rinki                   & 257,266        \\ 
			sawatimehera            & 239,962        \\ 
			chouhanshivraj          & 185,354        \\ 
			narendramodiforyou      & 181,271        \\ 
			anandranganathan        & 168,518        \\ 
			\bottomrule
			
		\end{tabular}
		\caption{Users with most number of likes. Ravi Shankar Prasad is the most liked user on Koo with 435,752 likes, while Republic takes the third place.}
		\label{tab:likes}
	}
\end{table}

\section{Discussion}

Koo's tagline, ``The Voices of India'', captures the essence of the platform, i.e., support for Indian languages, large Indian user base, and homegrown development. The recent surge in popularity of the platform, presence of multilingual content, and linguistic communities makes the study of Koo interesting. We release the first-ever dataset of Koo and characterize the platform based on the users, content posted, and the network. We show the formation of tight communities based on language, as well as the massive popularity of Indian politicians, news media agencies and government organizations on the network. We note that female users are more active, despite being present in smaller numbers. The higher presence of Hindi than English on Koo indicates a degree of success in promoting discourse in Indian languages. Kannada, Tamil, Telugu and Marathi also constitute a considerable portion of the content and activity on the platform. We observe a Koo vs Twitter rhetoric with hashtags such as ``\#koovstwitter`` and ``\#bantwitter`` trending. Koo is still in its nascent stages and is being developed to include more languages - Gujarati, Malayalam, Oriya, Punjabi, and Assamese. As Koo grows in size, it should build convenient mechanisms with which researchers can collect and work on their data, which can prove useful for research in social computing and Indian languages.

\section{Limitations and Future Work}
This paper performs an exploratory analysis of the characteristics of Koo, presenting a novel dataset and uncovering valuable insights. Although our dataset contains a substantial portion of the user-base of Koo, it does have the limitation that it leaves out singletons and isolated communities. For future work, this multilingual data can act as a corpus for research in Indian languages. A study of posting trends by the same users on Koo and Twitter could reveal if users use the two platforms for different purposes, or if their activity on one mirrors that on the other. Our dataset of corresponding Koo and Twitter user IDs makes this data collection convenient.  An analysis of who gains popularity on Koo, as well as the presence of bot accounts could reveal more about the inter-user interactions on the platform.

\section*{Acknowledgements}
The authors would like to thank the annotators for their help and contribution for creating the dataset. The authors would also like to thank Nidhi Goyal, Samiya Caur and Shivangi Singhal for their helpful reviews and comments on the initial draft versions.

\bibliographystyle{IEEEbib}  
\bibliography{template}

\end{document}